\documentclass[review]{elsarticle}

\usepackage{lineno,hyperref}
\modulolinenumbers[5]

\journal{Journal of \LaTeX\ Templates}

\biboptions{numbers,sort&compress}

\begin{document}

\begin{frontmatter}

\title{Intrinsic states of deformed $N=Z$ nuclei in a quartet  formalism}

\author[address1]{M. Sambataro\corref{correspondingauthor}}
\cortext[correspondingauthor]{Corresponding author}
\ead{michelangelo.sambataro@ct.infn.it}

\author[address2]{N. Sandulescu}
\ead{sandulescu@theory.nipne.ro}

\address[address1]{Istituto Nazionale di Fisica Nucleare - Sezione di Catania, Via S. Sofia 64, I-95123 Catania, Italy}
\address[address2]{National Institute of Physics and Nuclear Engineering, P.O. Box MG-6, Magurele, Bucharest, Romania}

\begin{abstract}
The band structure of $N=Z$ nuclei is constructed from intrinsic states defined in terms of quartets. The simplest of these states is a condensate of collective quartets with isospin $T=0$. The other intrinsic states are built by promoting one quartet of the condensate to an excited $T=0$ configuration. From these intrinsic states, by angular momentum projection, band structures are generated  that approximate well the experimental ones. The projected states also reproduce to a very good extent the spectra
resulting from configuration interaction calculations based on the same quartets forming the intrinsic states. These results show that the quartet-based intrinsic states provide the appropriate framework to understand in a simple and intuitive manner the emergence of band-like structures in $N=Z$ nuclei. 

\end{abstract}

\begin{keyword}
deformed $N=Z$ nuclei, quartets, intrinsic states
\end{keyword}

\end{frontmatter}


\section{Introduction}

The phenomenon of nuclear quarteting has a long history in nuclear structure physics. Starting from the early 60's, various studies have discussed the role played by quartets, namely four-body $\alpha$-like structures made by two protons and two neutrons, in the structure self-conjugate nuclei
\cite{soloviev,flowers,arima,faraggi,eichler,catara,hasegawa,senkov}. In spite of this long history, several unexplored and yet interesting aspects have come to light only in recent years. 

In Ref. \cite{sasa_so5_exact}, we have shown on an analytic basis that the exact $T=0$ seniority-zero eigenstates of the isovector pairing Hamiltonian,
in $N=Z$ nuclei and in the realistic case of nucleons moving on a set of non-degenerate orbitals, are built in terms $T=0$ quartets. Quartets had previously emerged from analogous exact treatments of isovector and isoscalar Hamiltonians in the case of degenerate orbitals \cite{dobes} as well as for a special form of the isovector-plus-isoscalar Hamiltonian \cite{dukelsky}. Quite important, in a recent past, 
it has been shown that a very simple and effective way of approximating the eigenstates of a generic $pn$ Hamiltonian in $N=Z$ systems consists in representing the ground state as a condensate of quartets \cite{qcm_t1,qcm_t0t1,qm_t1t0,qm_qcm_t0t1} and the associated excited states by simply promoting one quartet of this condensate to an excited $T=0$ configuration \cite{sasa_exc}. 

The phenomenon of nuclear quarteting is relevant not only for the case of $pn$ pairing Hamiltonians but also for
general effective interactions describing realistic $N=Z$ nuclei. This has been evidenced in some recent studies 
where a description of the low-lying states of some $N=Z$ has been provided in terms of ``static" quartets. These have been assumed to represent the low-lying states of the nearest $T=0$ one-quartet system and kept unmodified  for  all the nuclei in a major shell \cite{qm_prl,qm_pd,qm_odd}.

The static definition of the quartets just mentioned is clearly not the most appropriate one since it neglects
the effect of the Pauli principle when two or more of these quartets have to coexist in the same nucleus. Owing to that, in Ref. \cite{sasa_band}, we have developed a new approach based on ``dynamical" quartets which are extracted from special trial states built for the $N=Z$ nucleus under study. These trial states  carry a total isospin $T=0$ and an undefined angular momentum and have been assumed to represent ``intrinsic'' states of the $N=Z$ nucleus. Various types of trial states have been introduced. The simplest one is just a condensate of one $T=0$ quartet, this quartet being in turn a linear superposition of quartets with various $J$'s. This state has been defined as the ``ground'' intrinsic state. Other intrinsic states have been introduced, such as the ``$\beta$'' and the ``$\gamma$'' intrinsic states, which differ from the ground one only because one quartet of the condensate has been promoted to an excited configuration. The structure of these intrinsic states, and so of the associated quartets, has been fixed by minimizing the energy of the states. As a final step, in Ref. \cite{sasa_band}, the spectra of $N=Z$ nuclei have been generated by carrying out configuration-interaction (CI) calculations in spaces built in terms of these quartets. The comparison between exact spectra and those obtained with static and dynamical quartets has evidenced a clear improvement of the quality of the approximation in the latter case.
 
The most interesting result of Ref. \cite{sasa_band} has been the observation of a close correspondence between the sets of quartets generated by the various trial states and the occurrence of well defined band-like structures in the spectra of the $N=Z$ nuclei under study. 
The fundamental question addressed by the present study is whether interpreting these trial states 
as intrinsic states is well-founded. 
To the extent that this is so one expects that the same band-like structures resulting from the
CI calculations should be generated  by projecting states of good angular momentum from these intrinsic states.
Moreover, there being a number of possible intrinsic states, a rigorous criterion on which states to involve
in the calculations has to be settled. In this work we will explore these issues. As a result, a novel and simpler quartet-based framework for understanding the band-like structures in $N=Z$ nuclei will emerge from this study.

In Section 2, we will review the definition of the intrinsic states in the formalism of quartets as introduced in Ref. \cite{sasa_band} and outline the projection technique employed in this work. The criterion adopted for the choice of the intrinsic states will then be illustrated. In the same section we will compare spectra of $N=Z$ nuclei in the $sd$ and $pf$ shells obtained from the projection of these states with those resulting from CI and shell model (SM) calculations as well as with the experimental data. In Section 3, we will summarize the results and draw the conclusions.

\section{Formalism and applications}

We work in a spherically symmetric mean field and, using the standard notation, we introduce the label
$i\equiv \{n_i,l_i,j_i\}$ to denote the quantum numbers of a single-particle state. We define the  most general $T=0$ quartet creation operator as
\begin{equation}
q^+_{JM}=\sum_{i_1j_1J_1}\sum_{i_2j_2J_2}\sum_{T'}
q_{i_1j_1J_1,i_2j_2J_2,{T'}}
[[a^+_{i_1}a^+_{j_1}]^{J_1{T'}}[a^+_{i_2}a^+_{j_2}]^{J_2{T'}}]^{JT=0}_{M},
\label{1}
\end{equation}
where $a^+_{i\tau}$ creates either a proton or a neutron (depending on the isospin projection $\tau$) on the state
$i$ and $M$ stands for the projection of the total angular momentum $J$.
No restrictions on the intermediate couplings $J_1T'$ and $J_2T'$ are introduced and the amplitudes 
$q_{i_1j_1J_1,i_2j_2J_2,{T'}}$ are supposed to guarantee the normalization of the operator.
We shall focus on systems of $N_\pi$ protons and $N_\nu$ neutrons such that $N_\pi=N_\nu$
and  $N_\pi+N_\nu =4n$ ($n=2,3$) and assume axially symmetry of these systems.

As a starting point we introduce the state 
\begin{equation}
|\Theta_{g}\rangle = \mathcal{N}_g(Q^+_g)^n|0\rangle ,
\label{2}
\end{equation}
where
\begin{equation}
Q^+_g =\sum_J \alpha^{(g)}_J(q^+_g)_{J0},
\label{3}
\end{equation}
\begin{equation}
(q^+_g)_{J0}=\sum_{i_1j_1J_1}\sum_{i_2j_2J_2}\sum_{T'}
q^{(g)}_{i_1j_1J_1,i_2j_2J_2,{T'}}
[[a^+_{i_1}a^+_{j_1}]^{J_1{T'}}[a^+_{i_2}a^+_{j_2}]^{J_2{T'}}]^{JT=0}_{0}
\label{31}
\end{equation}
and $\mathcal{N}_g$ is a normalization factor. $|0\rangle$ is a vacuum state that will be defined below.
$|\Theta_{g}\rangle$ is thus a condensate of the quartet $Q^+_g$ which is in turn a linear superposition of the quartets $(q^+_g)_{J0}$ with $J$'s running over a set of values to be specified. In order to fix $Q^+_g$,
we minimize the energy of the state $|\Theta_{g}\rangle$ with respect to the coefficients $q^{(g)}_{i_1j_1J_1,i_2j_2J_2,{T'}}$ and $\alpha_{g,J}$.
The state (\ref{2}) will be referred to as the ``ground'' intrinsic state.

In addition to the intrinsic state just defined, we introduce a set of other states which are generated by promoting one
of the quartets $Q^+_g$ of $|\Theta_{g}\rangle$ to an excited $T=0$ configuration. These states have the general form
\begin{equation}
|\Theta_k\rangle = \mathcal{N}_k Q^\dag_k (Q^\dag_g)^{(n-1)}|0\rangle ,
\label{8}
\end{equation}
with
\begin{equation}
Q^\dag_k =\sum_J \alpha^{(k)}_J(q^\dag_k)_{Jk},
\label{9}
\end{equation}
\begin{equation}
(q^+_k)_{Jk}=\sum_{i_1j_1J_1}\sum_{i_2j_2J_2}\sum_{T'}
q^{(k)}_{i_1j_1J_1,i_2j_2J_2,{T'}}
[[a^+_{i_1}a^+_{j_1}]^{J_1{T'}}[a^+_{i_2}a^+_{j_2}]^{J_2{T'}}]^{JT=0}_{k}
\label{39}
\end{equation}
and where, also in this case, the angular momentum $J$  runs over a definite set of values to be specified.
Assuming that the quartet $Q^+_g$ has already been fixed, we construct the new quartet $Q^+_k$ by minimizing the energy of $|\Theta_k\rangle$ 
with respect to the coefficients $q^{(k)}_{i_1j_1J_1,i_2j_2J_2,{T'}}$ and $\alpha^{(k)}_J$ (under the constraint of orthogonality when various states with the same $k$ are involved). The states (\ref{8}) will be 
identified with the value of the quantum number $k$.

Once the intrinsic states (\ref{2}) and (\ref{8}) (and so the associate quartets) have been fixed, the spectrum 
of a $N=Z$ nucleus can be constructed by performing a CI calculation in the space built with the quartets which 
characterize these states. By working in the so-called $m$-scheme, this space if formed by the states
\begin{equation}
|\Psi^{(n)}_{\overline M},\{N_{JM}\}\rangle = \prod_{J,M\in{(-J,J)} }(q^+_{JM})^{N_{JM}}|0\rangle
\label{1a}
\end{equation}
under the conditions
\begin{equation}
\sum_{JM}N_{JM}=n,~~~~~~~\sum_{JM}MN_{JM}=\overline{M}.
\end{equation}
The calculation requires first the orthonormalization of the states (\ref{1a}) and then the diagonalization of the Hamiltonian in this basis for the various ${\overline M}$ (in order to identify the angular momentum of the states). This is the approach which has been followed in Ref. \cite{sasa_band} and, in the following, it will be referred to with the acronym QM (Quartet Model). In the present paper we will follow, instead, a different path. 
Both the states (\ref{2}) and (\ref{8}) have an undefined angular momentum.
In order to explore the validity of these states as proper intrinsic states of a $N=Z$ nucleus, we will construct the spectrum by projecting states of good angular momentum from them. The projection technique which has been used employs standard tensor coupling rules which do not deserve special explanations.
What is instead worth of being noticed is the fact that when projecting $m$ intrinsic states associated to a given nucleus one can generate up to $m$ states with the same angular momentum $J$. This projection does not guarantee neither that these states are orthogonal with each other nor that the Hamiltonian matrix is diagonal with respect to them. Thus, in these cases, a proper definition of the spectrum implies that we have first to build an orthonormal basis out of these projected states and therefore to diagonalize the Hamiltonian in such a basis. The maximum size of these basis in the calculations that we are going to present has been 6.
Calculations have been carried out in the $sd$ and $pf$ shells by adopting the USDB \cite{usdb} and KB3G \cite{kb3g} interactions, respectively.
The vacuum state $|0\rangle$ of the previous expressions stands for the nucleus $^{16}$O for $sd$ shell nuclei and for $^{40}$Ca for $pf$ shell nuclei.
 
The first problem that has to be faced in the approach just described is that of defining the most appropriate set of intrinsic states to involve in the calculations. In this work we have adopted the criterion of selecting the intrinsic states on the basis of their energy. In Fig. 1, we show the energies of the lowest intrinsic states (with the associate value of the quantum number $k$) in the cases of  $^{24}$Mg, $^{28}$Si and $^{48}$Cr. The values of the angular momentum $J$ entering the summations of Eqs. (\ref{3}) and (\ref{9}) have been restricted to $J=0,2,4$ for $|\Theta_{g}\rangle$ and $|\Theta_0\rangle$ while to $J=k,k+1,k+2$ for $|\Theta_k\rangle$ with $k\neq 0$. Only in the case of $^{48}$Cr an extra $J=6$ quartet has been added to $|\Theta_{g}\rangle$.
In all cases the lowest intrinsic state has been found to be the condensate (\ref{2}), followed by a $k$=0 or 2 state (\ref{8}). The calculations for the three nuclei quoted above have been done with the intrinsic states lying below the dashed-dotted lines shown in Fig. 1.

In Figs. 2-4, the low-lying states obtained within the projected approach  are compared with the experimental spectra and the
results of QM and SM calculations. For the experimental spectra we have shown only the states with certain angular momenta and parities. 
The numbers next to each level of the QM and SM spectra give the overlaps with the corresponding projected states while those at the bottom represent the ground state energies. An overall good agreement is found among all the spectra of Figs. 2-4. With reference to the projected cases, in particular, we notice that this result is all the more remarkable if one thinks that these spectra simply result from a projection and a subsequent diagonalization in very reduced spaces (only a few units). The effect of this limited space is quite evident in the case of  $^{28}$Si where one generates only three J=2 projected states and, in correspondence with the second of them, one finds two QM and three SM $J=2$ states. This is the only case with a no clear correspondence among projected, QM and SM states.

To end this section, it is worth mentioning that a simple SM calculation provides only a sequence of states. Associating them with specific band-like structures, such as ground, $\beta$ and $\gamma$-like bands,  requires additional analysis. In Figs. 2-4 we have split the SM states in groups of levels following the correspondence with the band-like structures generated by the intrinsic states and by the QM calculations. It is remarkable that these band-like structures reproduce well the experimental data.

\section{Summary and conclusions}

In this paper we have shown that the band structure of deformed $N=Z$ nuclei can be associated with intrinsic
states defined in terms of quartets.  The simplest of these states is just a condensate of a collective 
quartet with isospin $T=0$ and an undefined angular momentum. The other intrinsic states are built by promoting one  quartet of this condensate to an excited $T=0$ configuration. A criterion has been introduced to select the most appropriate set of these states. We have shown that the band structure of $^{24}$Mg, $^{28}$Si and $^{48}$Cr can be generated to a very good extent by projecting states of good angular momentum from the above intrinsic states. This fact demonstrates that the emergence of band-like structures in $N=Z$ nuclei can be simply understood in terms of quartet-based intrinsic states. These intrinsic states remind analogous states employed in the past to describe the band structure of deformed nuclei, the basic difference being that, in a description of $N=Z$ nuclei,  quartets have replaced the original collective pairs \cite{maglione, maglione2}.

As a concluding remark, we like to emphasize the interesting analogy between the formalism
presented in the present work, based on general two-body interactions of SM type, 
and the one employed in the case of the isovector-plus-isoscalar proton-neutron pairing interaction,
discussed in Ref. \cite{sasa_exc}. In the latter case, it was evidenced that a very accurate approximation of the ground and excited states could be provided, respectively, by a condensate of $T=0$,$J=0$ quartets (each built with isovector and isoscalar pairs) and by states obtained by promoting one quartet of this condensate to an excited $T=0$ configuration. The intrinsic states employed in the present work appear to be a generalization of these states in the case of deformed systems. The basic difference between the states of Ref. \cite{sasa_exc}  and the intrinsic states of this work is that while the former have a well defined angular momentum, the latter do not. Thus, in the present case, in order to generate the spectrum of a $N=Z$ nucleus it has been necessary to go through an additional step, namely to project states of good angular momentum from the intrinsic states.

\section*{Acknowledgements}
This work was supported by a grant of the Romanian Ministry of Research and Innovation, CNCS - UEFISCDI, project number PCE 160/2021, within PNCDI III.

\bibliography{biblio}

\begin{thebibliography}{10}
\expandafter\ifx\csname url\endcsname\relax
  \def\url#1{\texttt{#1}}\fi
\expandafter\ifx\csname urlprefix\endcsname\relax\def\urlprefix{URL }\fi
\expandafter\ifx\csname href\endcsname\relax
  \def\href#1#2{#2} \def\path#1{#1}\fi

\bibitem{soloviev}
V.~Soloviev, Nucl. Phys. 18 (1960) 161.

\bibitem{flowers}
B.~Flowers, M.~Vujicic, Nucl. Phys. 49 (1963) 586.

\bibitem{arima}
A.~Arima, V.~Gillet, J.~Ginocchio, Phys. Rev. Lett. 25 (1970) 1043.

\bibitem{faraggi}
H.~Faraggi, A.~Jaffrin, M.-C. Lemaire, M.~Mermaz, J.-C. Faivre, J.~Gastebois,
  B.~Harvey, J.-M. Loiseaux, A.~Papineau, Ann. of Phys. 66 (1971) 905.

\bibitem{eichler}
J.~Eichler, M.~Yamamura, Nucl. Phys. A 182 (1972) 33.

\bibitem{catara}
F.~Catara, J.~M. Gomez, Nucl. Phys. A 215 (1973) 85.

\bibitem{hasegawa}
M.~Hasegawa, S.~Tazaki, R.~Okamoto, Nucl. Phys. A 592 (1995) 45.

\bibitem{senkov}
R.~Senkov, V.~Zelevinsky, Phys. At. Nucl. 74 (2011) 1267.

\bibitem{sasa_so5_exact}
M.~Sambataro, N.~Sandulescu, J. Phys. G: Nucl. Part. Phys. 47 (2020) 045112.

\bibitem{dobes}
J.~Dobes, S.~Pittel, Phys. Rev. C 57 (1998) 688.

\bibitem{dukelsky}
J.~Dukelsky, V.~Gueorguiev, P.~V. Isacker, S.~Dimitrova, B.~Errea, S.~H. Lerma,
  Phys. Rev. Lett. 96 (2006) 072503.

\bibitem{qcm_t1}
N.~Sandulescu, D.~Negrea, J.~Dukelsky, C.~Johnson, Phys. Rev. C 85 (2012)
  061303(R).

\bibitem{qcm_t0t1}
N.~Sandulescu, D.~Negrea, D.~Gambacurta, Phys. Lett. B 751 (2015) 348.

\bibitem{qm_t1t0}
M.~Sambataro, N.~Sandulescu, C.~Johnson, Phys. Lett. B 740 (2015) 137.

\bibitem{qm_qcm_t0t1}
M.~Sambataro, N.~Sandulescu, Phys. Rev. C 93 (2016) 054320.

\bibitem{sasa_exc}
M.~Sambataro, N.~Sandulescu, Phys. Lett. B 820 (2021) 136476.

\bibitem{qm_prl}
M.~Sambataro, N.~Sandulescu, Phys. Rev. Lett. 115 (2015) 112501.

\bibitem{qm_pd}
M.~Sambataro, N.~Sandulescu, Phys. Rev. C 91 (2015) 064318.

\bibitem{qm_odd}
M.~Sambataro, N.~Sandulescu, Phys. Lett. B 763 (2016) 151.

\bibitem{sasa_band}
M.~Sambataro, N.~Sandulescu, Phys. Lett. B 827 (2022) 136987.

\bibitem{usdb}
B.~Brown, W.~Richter, Phys. Rev. C 74 (2006) 034315.

\bibitem{kb3g}
A.~Poves, G.~Martinez-Pinedo, Phys. Lett B 430 (1998) 203.

\bibitem{maglione}
E.~Maglione, F.~Catara, A.~Insolia, A.~Vitturi, Nucl. Phys. A 397 (1983) 102.

\bibitem{maglione2}
E.~Maglione, F.~Catara, A.~Insolia, A.~Vitturi, Nucl. Phys. A 411 (1983) 181.

\end{thebibliography}

\newpage
\begin{figure}
\begin{center}
\includegraphics*[scale=0.5,angle=-90]{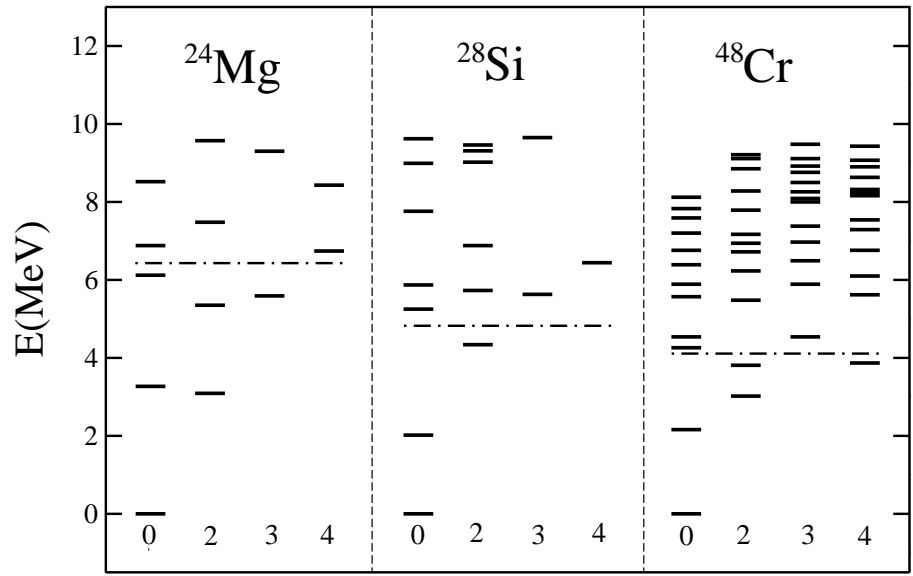}
\caption{Energies of the intrinsic states (\ref{2}) and (\ref{8}) for $^{24}$Mg, $^{28}$Si and $^{48}$Cr.
The structure of these states is specified in the text.
The numbers at the bottom represent the quantum number $k$ characterizing the intrinsic states. Only states with energies below the dashed-dotted lines have been included in the calculations.}
\end{center}
\end{figure}

\newpage
\begin{figure}
\begin{center}
\includegraphics*[scale=0.5,angle=-90]{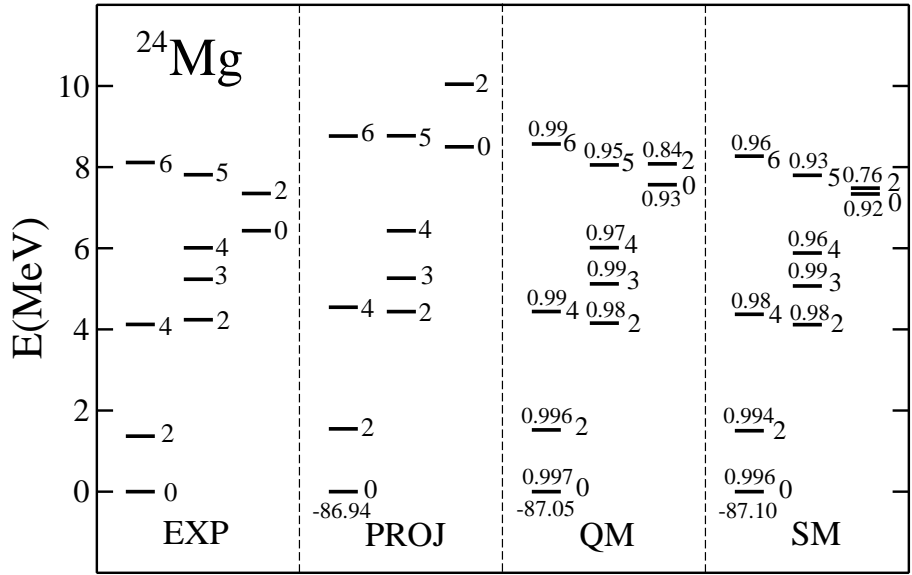}
\caption{$^{24}$Mg: experimental (EXP), projected (PROJ), QM and SM spectra. See text for details.}
\end{center}
\end{figure}

\newpage
\begin{figure}
\begin{center}
\includegraphics*[scale=0.5,angle=-90]{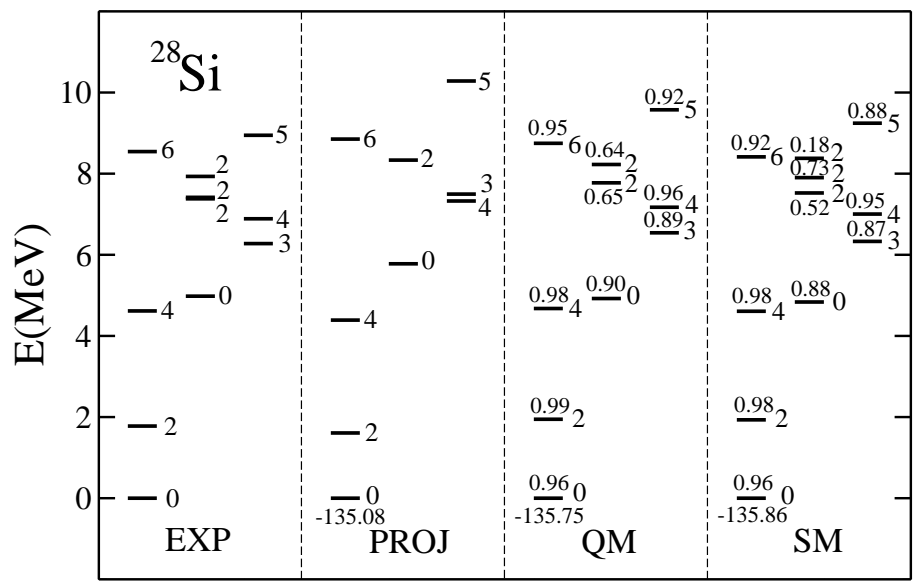}
\caption{As in Fig. 2 for $^{28}$Si}
\end{center}
\end{figure}

\newpage
\begin{figure}
\begin{center}
\includegraphics*[scale=0.5,angle=-90]{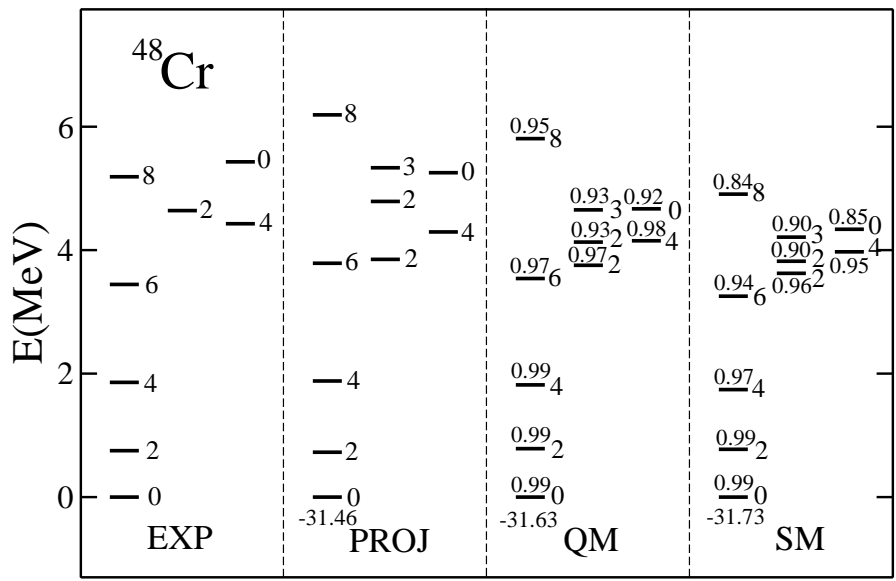}
\caption{As in Fig. 2 for $^{48}$Cr}
\end{center}
\end{figure}

\end{document}